\documentstyle[epsf]{article}
\def\epsfsize#1#2{\columnwidth}

\newcommand{\msolar}{M_{\odot}}

\newcommand{\mptr}{{\cal M}}
\newcommand{\fracparen}[2]{\left(\frac{#1}{#2}\right)}
\newcommand{\qddd}{\mbox{$\stackrel{\ldots}{\mbox{Q}}$}}

\newcommand{\oderiv}[2]{\frac{d #1}{d #2}}
\newcommand{\oderivn}[3]{\frac{d^{#3}\!#1}{d #2^{#3}}}

\newcommand{\oderivf}[2]{{d #1}\!/{d #2}}
\newcommand{\oderivnf}[3]{{d^{#3}\!#1}\!/{d #2^{#3}}}

\newcommand{\E}[1]{\times 10^{#1}}
\newcommand{\Eqref}[1]{Equation~(\ref{#1})}
\newcommand{\Figref}[1]{Fig.~\ref{#1}}

\newcommand{\hide}[1]{}\newcommand{\units}{\rm\;}
\newcommand{\BFS}{Bernard F. Schutz}
\newcommand{\Cardiff}{Department of Physics and Astronomy\\University of Wales College of Cardiff, Cardiff, U.K.}
\newcommand{\AEI}{Max Planck Institute for Gravitational Physics\\The Albert Einstein Institute\\Potsdam, Germany}

\begin{document}
\twocolumn
\title{Gravitational Radiation\thanks{%
To be published in the Encyclopedia of Astronomy and Astrophysics (Institute of Physics Publishing, Bristol, and Macmillan Publishers Ltd, London, 2000).}}
\author{\BFS\\ \AEI \\ and \\ \Cardiff }
\date{}
\maketitle


\subsection*{Gravitational wave astronomy}

Gravity is one of the fundamental forces of Nature, and it is the dominant force in most astronomical systems.  In common with all other phenomena, gravity must obey the principles of {\sc Special Relativity}.  In particular, gravitational forces must not be transmitted or communicated faster than light.  This means that when the gravitational field of an object changes, the changes ripple outwards through space and take a finite time to reach other objects.  These ripples are called {\em gravitational radiation} or {\em gravitational waves}.\footnote{They are also sometimes referred to as {\em gravity waves}, but since this term has a different meaning in meteorology and stellar hydrodynamics, we will avoid it here.  See {\sc Solar Interior: Helioseismology}.} 

In Einstein's theory of gravitation (see {\sc General Relativity and Gravitation}), as in many other modern theories of gravity (see {\sc Non-general Relativity Theories of Gravity}), gravitational waves travel at exactly the speed of light.  Different theories make different predictions, however, about details, such as their strength and polarization.  There is strong indirect observational evidence (see {\sc Binary Stars as a Probe of General Relativity, Hulse-Taylor Pulsar}) that gravitational waves follow the predictions of general relativity, and instruments now under construction are expected to make the first direct detections of them in the first years of the 21st century.  

These instruments and plans for future instruments in space are described in the article {\sc Gravitational Radiation Detection on Earth and in Space}.  Detectors must look for gravitational radiation from astronomical systems, because it is not possible to generate detectable levels of radiation in the laboratory.  It follows that gravitational wave detection is also a branch of observational astronomy.

The most striking aspect of gravitational waves is their weakness. A comparison with the energy in light will illustrate this. The human eye has no trouble sensing the light from the planet Jupiter: the amount of energy that passes through the iris of the eye is far more than the minimum the eye can detect.  Yet several times a week a gravitational wave, generated in a far distant galaxy, carries a similar amount of energy into the eye, and we don't notice it.  

The reason is that gravity is the weakest of the fundamental forces, and the disturbance created by even such an energetic wave is so tiny that no man-made instrument has so far registered it. While all the energy in the light from Jupiter that enters the eye is absorbed in the eye, the gravitational wave passes right through, leaving behind almost none of its energy.  All the matter in the present Universe is similarly transparent to gravitational waves.

Gravitational radiation is today one of the last unopened windows into the Universe.  There are at least five reasons motivating scientists to develop gravitational wave astronomy:
\begin{itemize}
\item The weakness with which gravitational waves interact with matter is a great advantage for astronomy.  It means that gravitational waves arrive unaffected by any intervening matter they may have encountered since being generated.  There is no significant scattering or absorption, although they will be deflected by a {\sc Gravitational Lens} in the same way as light.  Gravitational waves carry uncorrupted information even if they come from the most distant parts of the Universe or from its most hidden regions, like the interiors of {\sc Supernovae}.
\item Gravitational waves are emitted by the bulk motions of their sources, not by individual atoms or electrons, as is normally the case for electromagnetic waves.  They therefore carry a completely different kind of information about their sources from that which is normally available in observations of {\sc Binary Stars}, supernovae, and {\sc Neutron Stars}.  
\item Gravitational waves can be emitted by {\em black holes}, which are described in the article {\sc General Relativity and Gravitation}.  Indeed, gravitational waves provide only way to make {\em direct} observations of these objects.  Since there is now strong indirect evidence that giant black holes inhabit the centers of many (or even most) galaxies (see {\sc Supermassive black holes in AGN}), and since smaller ones are common in the Galaxy (see {\sc Black Hole Candidates in X-Ray Binaries}), there is great interest in making direct observations of them.
\item Gravitational waves can come from extraordinarily early in the history of the Universe.  The electromagnetic radiation from the Big Bang is called the {\sc Cosmic Microwave Background}.  Observations of it describe the Universe at it was about $10^5$ years after the Big Bang.  Studies of cosmological {\sc Nucleosynthesis} give information about what the Universe was like as little as 3 minutes after the Big Bang.  Gravitational waves, if they can be detected, would picture the Universe when it was only perhaps $10^{-24}$ seconds old, just at the end of {\sc Inflation}.  
\item Gravitational radiation is the last fundamental prediction of Einstein's general relativity that has not yet been directly verified.  If another theory of gravity is correct, then differences could in principle show up in the properties of gravitational waves, such as their polarization.  In principle, there must be a better theory of gravity, since general relativity is not a quantum theory, a deficiency that theoretical physicists today are working hard to remedy.  The majority belief today is that there should be a unified theory of the fundamental forces, in which gravitation is related to the other forces.  Evidence for the nature of this relation could show up in observations of gravitational waves, particularly those from the Big Bang.
\end{itemize}

These motivations and their implications are developed in the following sections.  Each section begins with an introduction to the physical ideas and then develops some of the mathematical details.  

\subsection*{The physics of gravitational radiation: weakness and strength}
The starting point for understanding gravitational radiation is Newtonian gravity.  The weakness of gravity is evident. If a child picks up a book, she defeats the cumulative gravitational pull of the entire planet Earth on the book.  The strength to do this comes from the chemical forces in her muscles, which come from electromagnetic interactions.  

In fact, the electromagnetic force between the electron and the proton in a hydrogen atom is $2\E{39}$ times bigger than the gravitational force between them.  The reason that gravity can nevertheless dominate on the cosmic scale is that opposite electrical charges cancel each other, while the gravitational forces of all the particles add.  

Another fact about gravity that was known to Newton is what is now called the {\em equivalence principle} (see {\sc General Relativity and Gravitation}). This is the principle that all bodies accelerate in the same way in a gravitational field, so that the trajectory that a {\em freely falling body} (a body influenced only by gravity) follows in a given gravitational field depends only on its starting position and velocity, not on what it is made of.

Imagine now a machine made in some way to detect gravitational waves.  Whatever the method of detection, a wave needs somehow to alter the internal state of the detector. If the wave carries a gravitational field that is completely uniform across the detector, then by the equivalence principle all of the parts of the detector will accelerate together, and its state will not change at all.  To detect a gravitational wave, the machine must measure the non-uniformities of the gravitational field across a detector.

These non-uniformities are called {\em tidal forces}, because they produce the stretching effects that raise tides on the Earth.  Gravitational waves are traveling tidal forces. 

Newton's theory of gravity had no gravitational waves.  For Newton, if a gravitational field changed in some way, that change took place instantaneously everywhere in space.  This is not a wave.   Let us consider what we mean by the term ``wave'' in ordinary language. 

Imagine a child's rubber duck floating in a bath tub half full of water. If a child presses down on the duck very gently, until is is nearly submerged, then the level of the water will rise everywhere in a nearly uniform way, and this is not called a wave.  If instead he drops the duck, then the disturbance rises around the base of the duck rapidly, moves away from it, and eventually reaches the walls.  This is a wave.  Wave motion requires a finite speed for the propagation of disturbances.   If the disturbance is very slow, as for the floating duck, then the wavelength is very long, and near the site of the disturbance the wave motion is not noticeable. We say we are in the ``near zone''.  But when we are more than a wavelength away, then we see waves, and this is the ``wave zone'' or ``far zone''.  For the dropped duck, we see that waves because their wavelengths are shorter than the size of the bath. In general relativity, the speed of gravity is the speed of light. Because of this finite speed, gravity must exhibit wave effects.

Many of Newton's contemporaries were unhappy that his theory of gravity was based on instantaneous ``action at a distance'', but Newton's theory fit the observational facts. If gravity had a finite speed of propagation, there was no evidence for it in the solar system.  Interestingly, the brilliant 18th century French mathematician and physicist Laplace tried out a variation on Newton's theory in which gravity was represented by something ``flowing'' out of its source with a finite speed.  He reasoned that a planet like the Earth, moving through this fluid of gravity, would experience friction and gradually spiral in towards the Sun.  

Laplace could show that the observational limits on this inspiral even in his day were so stringent that the speed of gravity in his model needed to be huge compared to the speed of light.  He did not find this result attractive and took the theory no further.  (Laplace also explored the notion of what we now call a  black hole, which for him was a region where gravity was strong enough to trap light.)

It is interesting that today, observations of the two  neutron stars in the binary system PSR1913+16 spiraling together as they orbit one another provides the most convincing evidence that gravitational waves exist and are as described in general relativity. (See below and the article {\sc Binary Stars as a Probe of General Relativity}.)  Laplace had the right effect, but the wrong theory.  This evidence is described in the next section.  

In general relativity, Einstein used the  principle of equivalence as the basis for a {\em geometrical} description of gravity.  In the four-dimensional world of space-time, the trajectory of a particle falling freely in a gravitational field is a certain fixed curve.  Its direction at any point depends on the velocity of the particle.  The equivalence principle implies that there is a preferred set of curves in space-time: at any point, pick any direction, and there is a unique curve in that direction that will be the trajectory of any particle starting with that velocity.  These trajectories are thus properties of space-time itself.  

Moreover, if there were no gravitational field, the trajectories would be simple straight lines.  Even in a gravitational field, a small freely falling particle does not ``feel'' any acceleration: its internal state is the same as if there were no gravity.  Therefore Einstein postulated that a gravitational field made space-time curved, and that the preferred trajectories were locally straight lines that simply changed direction as they moved through the curved space-time, in much the same way as a great circle on a sphere changes direction relative to other great circles as one goes along it.  For weak gravitational fields of slowly moving bodies, Einstein's theory reduces to Newton's in the first approximation.

For gravitational waves, one could make a very simple detector just by monitoring the distance between two nearby freely falling particles.  If they are genuinely free, then any changes in their separations would indicate the passage of a gravitational wave.  Because this measures a tidal effect, the bigger the separation of the particles, the bigger will be the change in their separation, at least for particles that are separated by less than a gravitational wavelength.  Most modern gravitational wave detectors are designed to be as big as cost and practicality allows.  These are described in the article {\sc Gravitational Radiation Detection on Earth and in Space}.

Although gravitational radiation is well understood in theoretical terms in 
general relativity, the complexity and non-linearity of Einstein's equations 
means that calculations are often difficult.  In the historical development 
of general relativity, between 1915 and the 1950's and 1960's, these 
mathematical difficulties created confusion over the physical nature of 
gravitational radiation, and in particular over whether they carried 
energy away from the source.  Improved mathematical techniques finally 
resolved the matter in favor of the simple physical picture presented 
here, but this picture would not be complete without the strong mathematical 
underpinning that now exists.

The question of energy in gravitational waves is still a delicate one.  
There is no question that waves carry energy (and momentum) away from 
their sources.  Nevertheless, it is not possible in general relativity to 
localize the energy in the radiation to regions smaller than about 
a wavelength.  Indeed the equivalence principle shows that ``point'' 
particles feel nothing, no matter how strong the wave.  The wave only 
acts by stretching space-time, producing a tidal distortion in the 
separations between particles (see the discussion of polarization below).  

For this reason, energy is localized only in regions, not at points.  It 
is nevertheless real energy: the nonlinearity of general relativity allows 
waves to create gravitation themselves.  Recent numerical simulations have 
shown that focussed gravitational waves can actually form black holes, 
trapping themselves.  If the waves are weak, they enter the focussing 
region and re-emerge.  If they are strong enough, they enter and never leave.

\subsubsection*{Gravitational waves in a quasi-Newtonian model}
It is possible to calculate the approximate size of the effect of a given gravitational wave by beginning with Newtonian gravity and adding waves to it. In Newtonian gravity the gravitational field produced by a mass $M$ at a distance $r$ is given by 
\begin{equation}\label{eqn:newtgrav}
\phi = -GM/r, 
\end{equation}
where $G$ is Newton's gravitational constant. 
The field of a gravitational wave must be a ripple on this, which means a small change that oscillates in space and time.  A suitable form for a {\em change} that propagates at the speed of light in the $z$-direction with an angular frequency $\omega$ is:
\begin{equation}\label{eqn:wavemodel}
\delta\phi = -\epsilon \frac{GM}{r}\sin[\omega(z/c-t)],
\end{equation}
where $\epsilon$ is a dimensionless number that would be expected to be small compared to 1.  Its size is the subject of the next main section. 

The field $\delta\phi$ produces an acceleration in the $z$-direction that depends on its $z$-derivative.  Both $1/r$ and the $\sin()$ term depend on z.  The derivative of $1/r$ will be proportional to $1/r^2$, which is how the acceleration falls off in Newton's theory (where $\phi$ is the only field).  But the derivative of the $\sin()$ term does not change the $1/r$; rather, it essentially just multiplies $\delta\phi$ by $\omega/c$.  At sufficiently large distances from the source, this term will dominate the $1/r^2$ term and the acceleration produced by the wave will be:
\begin{equation}
\delta a_z = -\epsilon \omega \frac{GM}{rc}\cos[\omega(z/c-t)].
\end{equation}
Note that this term would not be present in the $x$- and $y$-derivatives, so these components of the acceleration are much smaller in this quasi-Newtonian model of a gravitational wave.

\subsubsection*{Effect on a simple detector}
The tidal part of this acceleration, for a detector that has size $\ell$ in the $z$-direction, is to a first approximation
\begin{equation}\label{eqn:tidalforce}
\ell\cdot\oderiv{}{z}\delta a_z = \epsilon\omega^2 \ell \frac{GM}{rc^2} \sin[\omega(z/c-t)].
\end{equation}
If a detector consists of two freely falling particles with this relative acceleration, the equation of motion for their separation $\ell$ will be 
\begin{equation}\label{eqn:eom}
\oderivn{\ell}{t}{2} = \epsilon\omega^2 \ell \frac{GM}{rc^2} \sin[\omega(z/c-t)],
\end{equation}

The dimensionless coefficient $\epsilon(GM/rc^2)$ is typically very small. Even if  $\epsilon$ is of order 1, the other number is, with reasonable values for $M$ and $R$, 
\[\frac{GM}{rc^2} = 2.4\E{-21}\fracparen{M}{\msolar}\fracparen{r}{20\units Mpc}^{-1}.\]
The mass and distance scales here are those appropriate to a  neutron star in the nearest large cluster of galaxies, the Virgo Cluster. (The distance unit is based on the astronomers' parsec, denoted pc, which is about $3\times10^{16}$~m. The unit Mpc is a megaparsec.) It is believed that several neutron stars are formed in the Virgo Cluster each year in supernova explosions. Ever since the beginning of the development of gravitational wave detectors such events have been high on the list of possible sources of gravitational waves.

To solve \Eqref{eqn:eom}, one takes $r$ and $z$ as constants, and uses the
smallness of the right-hand-side, which implies that the changes in $\ell$ are tiny compared to $\ell$ itself. On the right-hand side one can therefore replace $\ell$ by $\ell_0$, the initial value of $\ell$, and then simply integrate twice in time to get (for an initial value  $\oderivf{\ell}{t} = 0$)  
\begin{equation} \label{eqn:effect}
\frac{\ell(t) - \ell_0}{\ell_0} = - \epsilon\frac{GM}{rc^2}\sin[\omega(z/c-t)].
\end{equation}
The right-hand-side of \Eqref{eqn:effect} is identical to that of \Eqref{eqn:wavemodel}.  This is an important conclusion which fits neatly with Einstein's geometrical conception of gravity: {\em the size of a gravitational wave gives directly the stretching of the distance between nearby free particles.}  It is conventional to call this $h/2$ and refer to $h$ as the  gravitational wave potential 
\begin{equation}\label{eqn:hdef}
h := 2\frac{\ell(t) - \ell_0}{\ell_0} = 2\delta\phi. 
\end{equation}
The amplitude of the oscillations of $h$ is 
\begin{equation}\label{eqn:hmax}
h \sim 2\epsilon\frac{GM}{rc^2}.
\end{equation}

It is evident from this that a detector must be able to measure changes in its own size that are smaller than one part in $10^{21}$ to have a reasonable chance of making astronomical observations.  The extraordinary smallness of this effect also explains why ordinary objects in the Universe are transparent to gravitational waves.  As the waves pass through them, they disturb them so little (parts per $10^{21}$ typically) that the transfer of energy to the object and any back-reaction effects of this on the wave are negligibly small.  

\subsubsection*{Energy flux carried by waves}
The {\em energy} in the waves can also be estimated from these equations and general physical principles. Quite generally, in classical field theories, the energy flux of a propagating sinusoidal plane wave is proportional to the square of the time-derivative of the fundamental field.  In electromagnetism, the Poynting flux is proportional to the square of the time-derivative of the vector potential.  

In general relativity, the flux is therefore proportional to the square of the time-derivative of $h(t)$.  The proportionality constant must be built only out of $c$, $G$, and pure numbers. To get the right units, it must be proportional to $c^3/G$; to get the pure number, a calculation in general relativity is required: $1/32\pi$ for a linearly polarized wave.  (Polarization is described later.)  This gives
\begin{eqnarray}
F_{gw}&=&\frac{1}{32\pi}\frac{c^3}{G}\fracparen{dh}{dt}^2 \label{eqn:flux} \\
 &=& 1.6\E{-5} \fracparen{f}{100 \units Hz}^2 \fracparen{h}{10^{-22}}^2 \units W\;m^{-2}, \nonumber
\end{eqnarray}
for a wave with frequency $f = \omega/2\pi$.  For comparison, reflected sunlight from Jupiter has a flux on Earth of $2.3\E{-7}\units W\;m^{-2}$, almost 100 times smaller than that of a gravitational wave with an amplitude of $10^{-22}$!

\subsubsection*{Deficiencies of the quasi-Newtonian model}
The calculation and equations in this section have been framed within a modified Newtonian model of gravity with a propagation speed of $c$, and one would expect some differences from general relativity.  The most important difference is in the {\em direction} in which the tidal forces act.  In the simple model, wave accelerations act in the $z$-direction, which was the direction of propagation of the wave.  This is called a {\em longitudinal wave}.  

In general relativity, gravitational waves are {\em transverse waves}: if the wave propagates in the $z$-direction then the tidal forces act only in the $xy$-plane.  We will discuss later the exact form that their action in this plane takes.  Remarkably, the rest of the formulas above are good approximations even in general relativity, provided $\epsilon$ is calculated correctly, as described in the next section.  

\subsection*{The emission of gravitational waves}
The previous section described the propagation of gravitational waves, their interaction with detectors, and the energy they carry.  This section deals with the strength with which waves are emitted by astronomical bodies.

In Newtonian gravity there is a fundamental theorem, proved by Newton, that the gravitational field outside a spherical body is not only spherical, but it is the same as that of a point mass located at the origin of the body.  It has the form given in \Eqref{eqn:newtgrav}.  In particular, the field is independent of the size of the body, as long as we consider only points outside it.  This is true even if the star pulsates in a spherical manner. 

This theorem is essentially the same in general relativity, and is known as {\em Birkhoff's Theorem}.  Outside a spherical body the field is the same as that of a black hole of the same mass as the body (the Schwarzschild metric), even if the body is pulsating spherically.  But if the pulsation is nonspherical, then the outside field will change.  In general relativity the changes generally propagate as a wave.  So gravitational waves will be emitted by nonspherical motions. 

In general the calculation of the emitted waves is extremely difficult, since the field equations of general relativity are a system of many coupled, nonlinear, partial differential equations.  But in four circumstances the emission mechanisms are understood in some detail:
\begin{itemize}
\item Small-amplitude pulsations of relativistic stars and  black holes.  Normally gravitational radiation carries away energy and damps pulsations away, but in rotating stars the opposite may happen: the radiated loss of angular momentum may allow the star to spin down to an energetically more favored state, in which case the perturbation will grow, at least until nonlinear effects intervene.  Discovered by S.\ Chandrasekhar and now called the {\em Chandrasekhar-Friedman-Schutz (CFS) instability}, it is thought to limit the rotation speed of {\em young}  neutron stars (see below).  Black holes also emit gravitational radiation when they are disturbed, {\em e.g.} by something falling into them, but they are not unstable: they always settle down into a steady state again.
\item Radiation from ``test'' objects orbiting  black holes.  If the mass of the object is small enough then the total gravitational field may be treated as a linear perturbation of the exactly known field of a black hole (the Schwarzschild or, with rotation, the Kerr solution).  These studies give insight into the general problem of gravitational radiation, and they also predict gravitational waveforms that might be observed by space-based observatories looking at compact stars falling into the giant black holes in the centers of galaxies. (See below and the article {\sc Gravitational Radiation Detection on Earth and in Space}.)  
\item Weak gravitational fields and slow motion.  Such weakly relativistic sources are studied in the {\em post-Newtonian} approximation, which includes higher-order corrections to Newtonian gravity from general relativity. This is analogous to the slow-motion multipole approximation that is so powerful in the study of electromagnetic radiation. Most realistic gravitational-wave sources can be studied to some approximation this way.
\item Collisions of black holes and neutron stars.  These events, which are expected to be observed by gravitational wave detectors (see below), must be modeled by solving the full set of Einstein equations on a powerful computer.  Techniques to do this are advancing rapidly, and simulations of realistic mergers of stars and black holes from in-spiraling orbits can be expected to yield useful results in the first years of the 21st century.
\end{itemize}

\subsubsection*{Quadrupole approximation}
The post-Newtonian approximation has so far been the most powerful of these methods, and it yields the most insight into the emission mechanisms.  Its fundamental result is the {\em quadrupole formula}, which gives the first approximation to the radiation emitted by a weakly relativistic system.  

The quadrupole formula is analogous to the dipole formula of electromagnetism.  In this language, {\em monopole} means spherical, which emits no radiation.  This is also true in electromagnetism, where it is linked to conservation of charge.  The ``monopole moment'' in electromagnetism is the total charge of a system, and since that does not change, there can be no spherical radiation.  

Again in electromagnetism, the dipole moment is defined as the integral
\[ d_i = \int \rho x_i d^3x,\]
where $\rho$ is the charge density and $x_i$ is a Cartesian coordinate.  If this integral is time-dependent, then the amplitude of the electromagnetic waves will be proportional to its first time-derivative $\oderivf{d_i}{t}$, and the radiated energy will be proportional (as we remarked earlier) to the square of the time derivative of this amplitude, {\em i.e.\ } to $\sum_i|\oderivnf{d_i}{t}{2}|^2$.  

In the post-Newtonian approximation to general relativity, the calculation goes remarkably similarly.  The monopole moment is now the total mass-energy, which is the dominant source of the gravitational field for non-relativistic bodies, and which is constant as long as the radiation is weak.  (Radiation will carry away energy, but in the post-Newtonian approximation that is a higher-order effect.)  The dipole moment is given by the same equation as above, but with $\rho$ interpreted as the density of mass-energy.  

However, here general relativity departs from electromagnetism.  The time-derivative of the dipole moment is, since the mass-energy is conserved, just the integral of the velocity $v_i$:
\begin{equation}\label{eqn:dipole}
\dot{d}_i = \int \rho v_i d^3x.
\end{equation}
But this is the total momentum in the system, and (to lowest order) this is {\em constant}.  Therefore, there is no {\em energy} radiated due to dipole effects in general relativity.  The gravitational field far from the source does contain a dipole piece if $\dot{d}_i$ is non-zero, but this is constant because it reflects the fact that the source has non-zero total momentum and is therefore moving through space.

To find genuine radiation in general relativity one must go one step beyond the dipole approximation to the quadrupole terms.  These are also studied in electromagnetism, and the analogy with relativity again is close.  The fundamental quantity is the spatial tensor (matrix) $Q_{jk}$, the second moment of the mass (or charge) distribution:
\begin{equation}\label{eqn:ibar}
Q_{jk}=\int\rho x_jx_k d^3x.
\end{equation}
A gravitational wave in general relativity is represented by a matrix $h_{jk}$ rather than a single scalar $h$, and its source (in the quadrupole approximation) is $Q_{jk}$.

As in electromagnetism, the amplitude of the radiation is proportional to the {\em second} time-derivative of $Q_{jk}$, and it falls off inversely with the distance $r$ from the source.  A factor of $G/c^4$ is needed in order to get a dimensionless amplitude $h$, and a factor of 2 to be consistent with the definition in \Eqref{eqn:hmax}. The result for $h_{jk}$ is:
\begin{equation}\label{eqn:hjk}
h_{jk} = \frac{2G}{rc^4}\oderivn{Q_{jk}}{t}{2}.
\end{equation}

General relativity describes waves with a matrix because gravity is geometry, and the effects of gravity are represented by the stretching of space-time.  This matrix contains that distortion information.  Here is the information about the transverse action of the waves that the quasi-Newtonian model of the last section did not get right. 

\subsubsection*{Simple estimates}
If the motion inside the source is highly non-spherical, then a typical component of $\oderivnf{Q_{jk}}{t}{2}$ will (from \Eqref{eqn:ibar}) have magnitude $Mv^2_{N.S.}$, where $v^2_{N.S.}$ is the non-spherical part of the squared velocity inside the source. So one way of approximating any component of \Eqref{eqn:hjk} is 
\begin{equation}\label{eqn:nonsph}
h\sim \frac{2GMv^2_{N.S.}}{rc^4}.
\end{equation}
Comparing this with \Eqref{eqn:hmax} we see that the ratio $\epsilon$ of the wave to the Newtonian potential is simply 
\[\epsilon \sim \frac{v^2_{N.S.}}{c^2}.\]
By the virial theorem for self-gravitating bodies, this will not be larger than 
\begin{equation}\label{eqn:epsilonandphi}
\epsilon <  \phi_{int}/c^2,
\end{equation}
where $\phi_{int}$ is the maximum value of the Newtonian gravitational potential {\em inside} the system.  This provides a convenient bound in practice.  It should not be taken to be more accurate than that.

For a neutron star source one has $\phi_{int} \sim 0.2 c^2$.  If the star is in the Virgo cluster, then the upper limit on the amplitude of the radiation from such a source is $5\E{-22}$.  {\em This has been the goal of detector development for decades, to make detectors that can observe waves at or below an amplitude of $10^{-21}$.}

\subsubsection*{Polarization of gravitational waves}
The matrix nature of the wave amplitude comes from general relativity and has no Newtonian analog.  In order to find the effect of the waves on the separation of two free particles (the idealized detector), one has to start with $h_{jk}$ as given by \Eqref{eqn:hjk} or by any other calculation, and then do three things:
\begin{enumerate}
\item Project the matrix $h_{jk}$ onto a plane perpendicular to the direction of travel of the wave.  In the simple case considered above, where the wave was traveling in the $z$-direction, this means leaving the components $\{h_{xx},\; h_{xy},\; h_{yy}\}$ alone and setting the remaining components to zero.  It is then a two-dimensional matrix in the transverse plane.
\item Remove the two-dimensional trace of the projected matrix.  Call the resulting matrix $h_{jk}^{TT}$, where {\em TT} stands for {\em Transverse-Traceless}.  In the example this means subtracting $(h_{xx}+h_{yy})/2$ from both $h_{xx}$ and $h_{yy}$. Then there are only two independent components left, $h_{xy}^{TT} = h_{yx}^{TT}$ and $h_{xx}^{TT} = -h_{yy}^{TT}$.
\item To find the change in the separation of two particles that have an initial separation given by the vector $\ell_k$, let the matrix $h_{jk}^{TT}$ act on it:
\begin{equation}\label{eqn:action}
\delta\ell_j = \sum_k h_{jk}^{TT}\ell_k.
\end{equation}
\end{enumerate}

It is clear that any longitudinal component of the separation $\ell_j$ between the particles is unaffected by the wave (in the example, this is the $z$-separation), and that there are two degrees of freedom (the two independent components of $h_{jk}^{TT}$) to move particles in the plane perpendicular to the propagation direction.  These two degrees of freedom are the two {\em polarizations} of the wave.  

\begin{figure}
\epsffile[20 40 576 343]{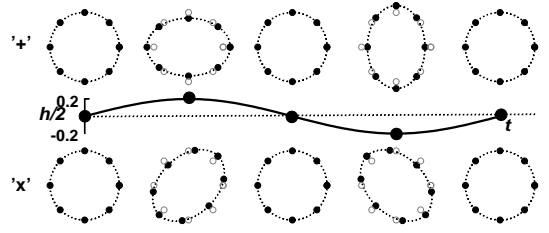}
\caption{Polarization of gravitational waves. The center line gives the wave as a function of time, with an amplitude of $h=0.2$, and the top and bottom lines show to scale the distortions produced by two polarizations with this amplitude.}\label{fig:poln}
\end{figure}

\Figref{fig:poln} shows the conventional definition of the two independent polarizations, from which any other can be made by superposition. What is shown is the effect of a wave on a ring of free particles in a plane transverse to the wave.  The first line shows a wave with $h_{xy}=0$, conventionally called the ``+'' polarization.  The bottom line shows a wave with $h_{xx} = 0$, the ``$\times$'' polarization.

\subsubsection*{Luminosity in gravitational waves}
The energy carried by the gravitational wave must be proportional to the square of the time-derivative of the wave amplitude, so it will depend on the sum of the squares of the components $\oderivnf{Q_{jk}}{t}{3}$.  The energy flux falls off as $1/r^2$, but when integrated over a sphere of radius $r$ to get the total luminosity, the dependence on $r$ goes away, as it should.  The luminosity contains a factor $G/c^5$ on dimensional grounds, and a further factor of $1/5$ comes from a careful calculation in general relativity.  The result is the gravitational wave luminosity in the quadrupole approximation:
\begin{equation}\label{eqn:luminosity}
L_{gw} = \frac{G}{5c^5}\left(\sum_{j,k}\qddd_{jk}\qddd_{jk} - \frac{1}{3}\qddd^2\right),
\end{equation}
where $Q$ is the trace of the matrix $Q_{jk}$.  Its squared third derivative must be subtracted in order to ensure that spherical motions do not radiate.

This equation will be used in the next section to estimate the back-reaction effect on a system that emits gravitational radiation.

\subsection*{Emission estimates}
Until observations of gravitational waves are successfully made, one can only make intelligent guesses about most of the sources that will be seen.  There are many that could be strong enough to be seen by the early detectors: binary stars, supernova explosions, neutron stars, the early Universe.  The estimates in this section are accurate only to within factors of order 1.  The estimates correctly show how the important observables scale with the properties of the systems.

\subsubsection*{Man-made gravitational waves}
One source can be ruled out: man-made gravitational radiation.  Imagine creating a wave generator with the following extreme properties.  It consists of two masses of $10^3$~kg each (a small car) at opposite ends of a beam 10~m long.  At its center the beam pivots about an axis.  This centrifuge rotates 10 times per second. All the velocity is non-spherical, so $v^2_{N.S.}$ in \Eqref{eqn:nonsph} is about $10^5\units m^2\;s^{-2}$.  The frequency of the waves will actually be 20~Hz, since the mass distribution of the system is periodic with a period of 0.05~s, only half the rotation period.  The wavelength of the waves will therefore be $1.5\E7$~M, about the diameter of the earth.  In order to detect gravitational waves, not near-zone Newtonian gravity, the detector must be at least one wavelength from the source.  Then the amplitude $h$ can be deduced from \Eqref{eqn:nonsph}: $h \sim 5\E{-43}$.  This is far too small to contemplate detecting!

\subsubsection*{Radiation from a spinning neutron star}
Some likely gravitational wave sources behave like the centrifuge, only on a grander scale.  Suppose a  neutron star of radius $R$ spins with a frequency $f$ and has an irregularity, a bump of mass $m$ on its otherwise axially symmetric shape.  Then the bump will emit gravitational radiation (again at frequency $2f$ because it spins about its center of mass, so it actually has mass excesses on two sides of the star), and the non-spherical velocity will be just $v_{N.S.} = 2\pi Rf$.  The radiation amplitude will be, from \Eqref{eqn:nonsph}, 
\begin{equation}\label{eqn:bump}
h_{bump}\sim 2(2\pi Rf/c)^2Gm/rc^2,
\end{equation}
and the luminosity, from \Eqref{eqn:luminosity} (assuming roughly 4 comparable components of $Q_{jk}$ contribute to the sum),
\[L_{bump}\sim (G/5c^5)(2\pi f)^6m^2R^4.\]
The radiated energy would presumably come from the rotational energy of the star. This would lead to a spindown of the star on a timescale
\[t_{spindown} = \frac{1}{2}mv^2/L_{bump} 
\sim \frac{5}{4\pi}f^{-1}\fracparen{Gm}{Rc^2}^{-1}\fracparen{v}{c}^{-3}.\]
It is felt that neutron star crusts are not strong enough to support asymmetries with a mass of more than about $m\sim10^{-5}\msolar$, and from this one can estimate the likelihood that the observed spindown timescales of {\sc Pulsars} are due to gravitational radiation.  In most cases, it seems that gravitational wave losses cannot be the main spindown mechanism.  

But lower levels of radiation would still be observable by detectors under construction, and this may be coming from a number of stars.  In particular, there is a class of neutron stars in {\sc X-Ray Binary Stars}.  They are accreting, and it is possible that accretion will create some kind of mass asymmetry or else lead to a rotational instability of the CFS type in the r-modes (see below).  In either case, the stars could turn out be long-lived sources of gravitational waves.  

\subsubsection*{Radiation from a binary star system}
Another ``centrifuge'' is a binary star system. Two stars of the same mass $M$ in a circular orbit of radius $R$ have $v^2_{N.S.} = GM/4R$.  The gravitational-wave amplitude from \Eqref{eqn:nonsph} can then be written
\begin{equation}\label{eqn:binary}
h_{binary} \sim  \frac{1}{2}\frac{GM}{rc^2}\frac{GM}{Rc^2}.
\end{equation}
Compare this to the implications of putting \Eqref{eqn:epsilonandphi} into \Eqref{eqn:hmax}.  

The gravitational-wave luminosity of such a system is, by a calculation analogous to that for bumps on neutron stars, 
\[L_{binary} \sim \frac{1}{80}\frac{c^5}{G}\fracparen{GM}{Rc^2}^5.\]
In this equation there appears the important constant $c^5/G = 3.6\E{52}$~W, 
a number with the dimensions of luminosity built only from fundamental constants.  By comparison, the luminosity of the Sun is only $3.8\E{26}$~W.  Close binaries can therefore radiate more energy in gravitational waves than in light. 

The radiation of energy by the orbital motion causes the orbit to shrink. The shrinking will make any observed gravitational waves increase in frequency
with time.  This is called a {\em chirp}. The timescale for this is
\begin{equation}\label{eqn:chirp}
t_{chirp} = Mv^2/L_{binary} \sim \frac{20GM}{c^3}\fracparen{GM}{Rc^2}^{-4}.
\end{equation}

\subsubsection*{The binary pulsar system -- verifying gravitational waves}
This orbital shrinking has already been observed in the {\sc Hulse-Taylor Pulsar} system, containing the radio pulsar PSR1913+16 and an unseen neutron star in a binary orbit.  Discovered in 1974 by R Hulse and J Taylor, it has established that gravitational radiation is correctly described by general relativity.  For their discovery, Hulse and Taylor received the 1993 Nobel Prize for Physics.

The key to the importance of this binary system is that all of the important parameters of the system can be measured before one takes account of the orbital shrinking.  This is because a number of post-Newtonian effects on the arrival time of pulses at the Earth, such as the precession of the position of the periastron and the time-dependent gravitational redshift of the pulsar period as it approaches and recedes from its companion, are measured in this system. They fully determine the masses and separation of the stars and the inclination and eccentricity of their orbit.  From these numbers, without any free parameters, it is possible to compute the shrinking timescale predicted by general relativity.  The observed rate matches the predicted rate to within the observational errors of less than 1\%.

The stars are in an eccentric orbit ($e = 0.615$) and both have masses of $1.4\msolar$.  The orbital semimajor axis is about $7\E8$~m.  \Eqref{eqn:chirp} assumes a circular orbit and gives a shrinking timescale of $6\E{10}$~y.  This is an overestimate, however, partly because it is in any case a rough approximation, and partly because the timescale is very sensitive to eccentricity.  With the observed eccentricity, a careful calculation shows that the expected shrinking timescale is around $4\E8$~y, consistent with observations.

\subsubsection*{Chirping binaries}
For a circular equal-mass binary, the orbital shrinking timescale and the frequency of the orbit determine both $M$ and $R$.  If in addition a gravitational wave detector measures the wave amplitude $h_{binary}$, then the {\em distance} $r$ to the binary system can be determined.
  
Remarkably, this conclusion holds even for binaries with unequal masses.  In such a case, the measurable mass is the {\em  chirp mass} of the binary, defined as $\mptr:=\mu^{3/5}M_T^{2/5}$, where $\mu$ is the reduced mass of the binary system and $M_T$ its total mass.  Then the distance $r$ is still measurable from the chirp rate, frequency, and amplitude.   In other words, {\em a chirping binary is a standard candle in astronomy.}  Post-Newtonian corrections to the orbit, if observed in the waveform, can determine the individual masses of the stars and even their spins.

\subsection*{Recognizing weak signals}
For ground-based detectors, all expected signals have amplitudes that are close to or even below the instrumental noise level in the detector output.  Such signals can nevertheless be detected with confidence if their waveform matches an expected waveform.  The pattern recognition technique that will be used by detector scientists is called {\em matched filtering}.  

Matched filtering works by multiplying the output of the detector by a function of time (called the {\em template}) that represents an expected waveform, and summing (integrating) the result.  If there is a signal matching the waveform buried in the noise then the output of the filter will be higher than expected for pure noise. 

A simple example of such a filter is the Fourier transform, which is a matched filter for a constant-frequency signal.  The noise power in the data stream is spread out over the spectrum, while the power in the signal is concentrated in a single frequency.  This makes the signal easier to recognize.  The improvement of the signal-to-noise ratio for the {\em amplitude} of the signal is proportional to the square-root of the number of cycles of the wave contained in the data.  This is well-known for the Fourier transform, and it is generally true for matched filtering.  

Matched filtering can make big demands on computation, for several reasons. First, the arrival time of a short-duration signal is generally not known, so the template has to be multiplied into the data stream at each distinguishable arrival time.  This is then a correlation of the template with the data stream. Normally this is done efficiently using Fast Fourier Transform methods.  

Second, the expected signal usually depends on a number of unknown parameters.  For example, the radiation from a binary system depends on the chirp mass $\mptr$, and it might arrive with an arbitrary phase.  Therefore, many related templates must be separately applied to the data to cover the whole family of signals.

Third, matched filtering enhances the signal only if the template stays in phase with the signal for the whole data set.  If they go out of phase, the method begins to reduce the signal-to-noise ratio.  For long-duration signals, such as for low-mass neutron-star coalescing binaries or continuous-wave signals from neutron stars (see below), this requires the analysis of large data sets, and often forces the introduction of additional parameters to allow for small effects that can make the signal drift out of phase with the template.  It also means that the method works well only if there is a good prediction of the form of the signal.  

Because the first signals will be weak, matched filtering will be used wherever possible.  As a simple rule of thumb, the detectability of a signal depends on its {\em effective} amplitude $h_{\mbox{\footnotesize \em eff}}$, defined as 
\begin{equation}\label{eqn:effective}
h_{\mbox{\footnotesize \em eff}} = h \sqrt{N_{cycles}},
\end{equation}
where $N_{cycles}$ is the number of cycles in the waveform that are matched by the template.  

For example, the effective amplitude of the radiation from a bump on a neutron star (\Eqref{eqn:bump})will be $h_{bump}\sqrt{2fT_{obs}}$, where $T_{obs}$ is the observation time.  In order to detect this radiation, detectors may need to observe for long periods, say 4 months, during which they accumulate billions of cycles of the waveform.  During this time, the star may spin down by a detectable amount, and the motion of the Earth introduces large changes in the apparent frequency of the signal, so matched filtering needs to be done with care and precision.

Another example is a binary system followed to coalescence, {\em i.e.\ } where the chirp time in \Eqref{eqn:chirp} is less than the observing time.  For neutron-star binaries observed by ground-based detectors this will always be the case (see the next section), so the effective amplitude is roughly 
\begin{equation}\label{eqn:chirpeffective}
h_{chirp} \sim h_{bin}\sqrt{f_{gw}t_{chirp}} \sim  \frac{GM}{rc^2}\fracparen{GM}{Rc^2}^{-1/4},
\end{equation}
where for $f_{gw}$ one must use twice the orbital frequency $\sqrt{GM/R^3}/4\pi$.  This may seem a puzzling result, because it says that the effective amplitude of the signal gets {\em smaller} as the stars get closer.  But this just means that the signal will be more detectable if it is picked up earlier, since \Eqref{eqn:chirpeffective} assumes that the signal is followed right to coalescence. If one picks up the signal at earlier times, then there are more cycles of the waveform to filter for, and this naturally gives a better signal-to-noise ratio.  This gives an advantage to detectors that can operate at lower frequencies. This has been an important consideration in the design of modern detectors. (See {\sc Gravitational Radiation Detection on Earth and in Space}.)

In general, the sensitivity of detectors will be limited not just by detector technology, but also by the duration of the observation, the quality of the signal predictions, and even by the availability of computer processing power for the data analysis.

\subsection*{Astronomical sources of gravitational waves}
\subsubsection*{Estimating the frequency}
The signals for which the best waveform predictions are available have narrowly defined frequencies. In some cases the frequency is dominated by an existing motion, such as the spin of a pulsar.  But in most cases the frequency will be related to the {\em natural frequency} for a self-gravitating body, defined as
\begin{equation}\label{eqn:natfreq}
f_0 = \sqrt{G\bar{\rho}/4\pi},
\end{equation}
where $\bar{\rho}$ is the mean density of mass-energy in the source.  This is of the same order as the binary orbital frequency and the fundamental pulsation frequency of the body.  

The frequency is determined by the size $R$ and mass $M$ of the source, taking $\bar{\rho} = 3M/4\pi R^3$.  For a  neutron star of mass $1.4\msolar$ and radius 10~km, the natural frequency is $f_0 = 1.9$~kHz.  For a black hole of mass $10\msolar$ and radius $2GM/c^2 = 30$~km, it is $f_0 = 1$~kHz.  And for a large black hole of mass $2.5\E6\msolar$, such as the one at the center of our Galaxy, this goes down in inverse proportion to the mass to $f_0 = 4$~mHz.  

\begin{figure}
\epsffile[120 300 460 550]{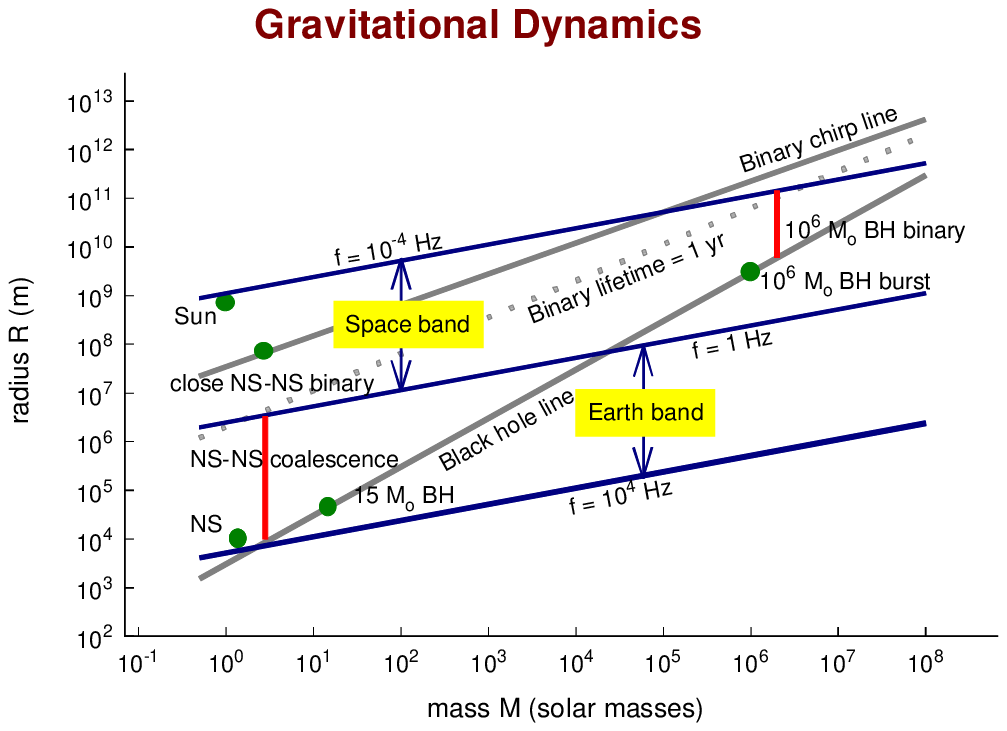}
\caption{Mass-radius plot for gravitational wave sources.}
\label{fig:freq}
\end{figure}

\Figref{fig:freq} shows the mass-radius diagram for likely sources of gravitational waves.  Three lines of constant natural frequency are plotted: $f_0 = 10^4$~Hz, $f_0 = 1$~Hz, and $f_0 = 10^{-4}$~Hz.  These are interesting frequencies from the point of view of observing techniques: gravitational waves between 1 and $10^4$~Hz are accessible to ground-based detectors, while lower frequencies are observable only from space. (See the article {\sc Gravitational Radiaton Detection on Earth and in Space}.)  Also shown is the line marking the black-hole boundary.  This has the equation $R=2GM/c^2$.  There are no objects below this line.  This line cuts through the ground-based frequency band in such a way as to restrict ground-based instruments to looking at stellar-mass objects.  Nothing over a mass of about $10^4\msolar$ can radiate above 1~Hz.  

A number of typical relativistic objects are placed in the diagram: a neutron star, a binary pair of neutron stars that spirals together as they orbit, some black holes.  Two other interesting lines are drawn.  The lower (dashed) line is the 1-year coalescence line, where the orbital shrinking timescale in \Eqref{eqn:chirp} is less than one year.  The upper (solid) line is the 1-year chirp line: if a binary lies below this line then its orbit will shrink enough to make its orbital frequency increase by a measurable amount in one year.  (In a one-year observation one can in principle measure changes in frequency of $1\units yr^{-1}$, or $3\E{-8}$~Hz.)  

It is clear from the figure that {\em any binary system that is observed from the ground will coalesce within an observing time of one year.}  Since pulsar statistics suggest that this happens less than once every $10^5$ years in our Galaxy, ground-based detectors must be able to register these events in a volume of space containing at least $10^6$ galaxies in order to have a hope of seeing occasional coalescences.  When detectors reach this sensitivity (sometime in the first decade of the 21st century), then astronomers will be able to use the observed chirping binaries as standard candles to measure distance scales in the Universe.

\subsubsection*{Radiation from neutron-star normal modes}
In \Figref{fig:freq} there is a dot for the typical  neutron-star.  The corresponding frequency is the fundamental vibrational frequency of such an object.  In fact, neutron stars have a rich spectrum of non-radial normal modes, which fall into several families: f-, g-, p-, w-, and r-modes have all been studied.  If their gravitational wave emissions can be detected, then the details of their spectra would be a sensitive probe of their structure and of the {\sc Equation of State of Neutron Stars}, in much the same way that {\sc Helioseismology} probes the interior of the Sun.  This is a challenge to ground-based detectors, which cannot yet make sensitive observations as high as 10~kHz.

\subsubsection*{Radiation from gravitational collapse}
The event that forms a neutron star is the gravitational collapse that produces in a supernova.  It is difficult to predict the waveform or amplitude expected from this event, because we have no observational evidence about how nonspherical the collapse event might be in a typical supernova: the collapse is hidden deep within the star.  So we can only guess. For example, a gravitational wave burst might be broad-band, centered on 1~kHz, or it might be a few cycles of radiation at a frequency anywhere between 100~Hz and 10~kHz, chirping up or down.  The amplitude could be large, in which a good fraction of the energy released by the collapse is radiated in gravitational waves, or it could be negligibly small. It is indeed ironic that, although detecting supernovae was the initial goal of detector development when it started 4 decades ago, little more is known today about what to expect than scientists knew then. 

\subsubsection*{Radiation from r-modes}
Hot neutron stars that rotate faster than about 100-200~Hz appear to be unstable to the emission of gravitational radiation through amplification of their r-modes by the CFS mechanism.  In stars colder than about $10^8$~K, viscosity may be strong enough to damp out this instability.  This instability may explain why only old, recycled, cold pulsars are seen at higher rotation rates.  It also suggests that the formation of a rapidly rotating neutron star may be followed by a period of steady gravitational radiation as the star emits angular momentum and spins down to its stability limit.  If as few as 10\% of all the neutron stars formed since {\sc Star Formation} began (at a redshift of perhaps 4) went through such a spindown, then they may have produced a detectable random background of gravitational radiation.

Interestingly, the r-modes are disturbances primarily of the fluid {\em velocity}; they have little density perturbation.  Their name comes from their similarity to the {\em R}ossby waves of oceanography.  The gravitational radiation they emit is not primarily mass-quadrupole (as in \Eqref{eqn:hjk}), but rather mass-current-quadrupole, the analog of magnetic quadrupole radiation in electromagnetism.  This is the wave counterpart of what is called {\em gravitomagnetism}, which is responsible for the {\em Lense-Thirring effect}: an extra precession of a spinning gyroscope as it orbits a rotating body like the Earth caused by the spin-spin gravitational coupling of the gyroscope to the Earth.
 
\subsection*{Black holes and gravitational waves}
Black holes are regions of spacetime within which everything is trapped: light cannot escape, nor can anything else that moves slower than light.  The boundary of this region is called the {\em event horizon}.  This boundary is a dynamical surface.  If any mass-energy falls into the hole, the area of the horizon increases.  In addition, the horizon will generally wobble when this happens.  These wobbles settle down quickly, emitting gravitational waves, and leaving a smooth (and slightly larger) horizon afterwards. 

Undisturbed black holes are time-independent and smooth.  In fact, according to general relativity the external gravitational field of such a black hole and the size and shape of its horizon are fully determined by only three numbers: the total mass, electric charge and angular momentum of the black hole.  This black-hole uniqueness theorem is remarkable, considering how much variety there can be in the material that collapsed to form the black hole and that may have subsequently fallen in.  

Observations of the gravitational waves emitted by a wobbling horizon or by a particle in orbit around a black hole have the potential to test the uniqueness theorem and thereby to verify the predictions of general relativity about the strongest possible gravitational fields.  

Astronomers now recognize that there is an abundance of black holes in the universe.  Observations of various kinds have located black holes in  X-ray binary systems in the Galaxy and in the centers of galaxies.  

These two classes of black holes have very different masses.  Stellar black holes typically have masses of around $10\msolar$, and are thought to have been formed by the gravitational collapse of the center of a large, evolved {\sc Red Giant Star}, perhaps in a supernova explosion.  Massive black holes in galactic centers seem to have masses between $10^6$ and $10^{10}\msolar$, but their history and method of formation are not yet understood.

Both kinds of black hole can radiate gravitational waves.  According to \Figref{fig:freq}, stellar black hole radiation will be in the ground-based frequency range, while galactic holes are detectable only from space. The radiation from a black hole typically is strongly damped, lasting only a few cycles about the frequency, which for a spherical black hole is given by \Eqref{eqn:natfreq} with $R=2GM/c^2$:
\[f_{BH}\sim 10 \fracparen{M}{\msolar}^{-1}\rm\; Hz.\]

\subsubsection*{Stellar-mass black holes}
Radiation from stellar black holes is expected mainly from coalescing binary systems, when one or both of the components is a black hole.  Although such systems are thought to be rarer than systems of two neutron stars, the larger mass of the black hole makes the system visible from a greater distance.  By measuring the chirp mass (as discussed above) observers will recognize that they have a black-hole system.  It is very possible that the first observations of binaries by interferometers will be of black holes.

When a two-black-hole binary coalesces, there should be a burst of gravitational radiation that will depend in detail on the masses and spins of the objects.  Numerical simulations of such events will be needed to interpret this signal, and possibly even to extract it from the instrumental noise of the detector.  The research field of {\em numerical relativity} is making rapid progress, and it can be expected to produce informative simulations in the first few years of the 21st century, using the largest and fastest computers available at that time.  

\subsubsection*{Massive and supermassive black holes}
Gravitational radiation is expected from supermassive black holes in two ways.  In one scenario, two massive black holes spiral together in a much more powerful version of the coalescence we have just discussed.  The frequency is much lower, but the amplitude is higher.  \Eqref{eqn:chirpeffective} implies that the effective signal amplitude is almost linear in the masses of the holes, so that a signal from two $10^6\msolar$ black holes will have an amplitude $10^5$ times bigger than the signal from two $10\msolar$ holes at the same distance.  Even allowing for differences in technology, space-based detectors will be able to study such events with a very high signal-to-noise ratio no matter where in the universe they occur.  

Observations of coalescing massive black-hole binaries will therefore provide strong tests of the validity of general relativity in the regime of strong gravitational fields, provided that numerical simulations can match the accuracy of the observations by that time.  

The event rate for such coalescences is not easy to predict: it could be zero, but it may be large.  It seems that the central core of most galaxies may contain a black hole of at least $10^6\msolar$.  This is known to be true for our galaxy and for a number of others nearby.  Supermassive black holes (up to a few times $10^9\msolar$) are believed to power {\sc Quasi-Stellar Objects} and {\sc Active Galaxies}.  There is some evidence that the mass of the central black hole is proportional to the mass of the core of the host galaxy.

If black holes are formed with their galaxies, in a single spherical gravitational collapse event, and if nothing happens to them after that, then coalescences will never be seen.  But it is believed that {\sc Galaxy Formation} probably occurred through the merger of smaller units, sub-galaxies of masses upwards of $10^6\msolar$.  If these units had their own black holes, then the mergers would have resulted in the coalescence of many of the holes on a timescale shorter than the present age of the universe.  This would give an event rate of several per year.  If the supermassive black holes were formed from smaller holes in a hierarchical merger scenario, then the event rate could be hundreds or thousands per year.  It is likely that only space-based observations of gravitational waves will answer these questions.

A second scenario for the production of radiation by massive black holes is the swallowing of a stellar-mass black hole or a neutron star by the large hole.  Massive black holes exist in the middle of dense star clusters.  The tidal disruption of main-sequence or giant stars that stray too close to the hole is thought to provide the gas that powers the quasar phenomenon.  These clusters will also contain a good number of neutron stars and stellar-mass black holes.  They are too compact to be disrupted by the hole even if they fall directly into it.  

Such captures therefore emit a gravitational wave signal that may be approximated by studying the motion of a ``point mass'' near a black hole.  It will again emit a chirp of radiation, but in this case the orbit may be very eccentric.  The details of the waveform encode information about the geometry of space-time near the hole.  In particular, it may be possible to measure the mass and spin of the hole and thereby to test the uniqueness theorem for black holes.  The event rate is not very dependent on the details of galaxy formation, and is probably high enough for many detections per year from a space-based detector.

\subsection*{Gravitational waves from the Big Bang}
Gravitational waves have traveled almost unimpeded through the universe since they were generated at times as early as $10^{-24}$~s after the Big Bang.  Observing them would provide important constraints on theories of {\sc Inflation} and high-energy physics.

Inflation is an attractive scenario for the early universe because it makes the large-scale homogeneity of the universe easier to understand.  It also provides a mechanism for producing initial density perturbations large enough to evolve into galaxies as the universe expands.  These perturbations are accompanied by gravitational-field perturbations that travel through the universe, redshifting in the same way that photons do.  Today these perturbations should form a random background of gravitational radiation.

The perturbations arise by parametric amplification of quantum fluctuations in the gravitational wave field that existed before inflation began.  The huge expansion associated with inflation puts energy into these fluctuations, converting them into real gravitational waves with classical amplitudes. 

If inflation did not occur, then the perturbations that led to galaxies must have arisen in some other way, and it is possible that this alternative mechanism also produced gravitational waves.  One candidate is cosmic defects, including {\em cosmic strings} and cosmic texture. (See {\sc Topological Defects in Cosmology}.)  Although observations at present seem to rule cosmic defects out as a candidate for galaxy formation, cosmic strings may nevertheless have produced observable gravitational waves.  

If inflation did not occur, there could also be a thermal background of gravitational waves at a temperature similar to that of the cosmological microwave background, but this radiation would have such a high frequency that it would not be detectable by any known or proposed technique.

The random background will be detectable as a noise in the detector that competes with instrumental noise.  In a single detector, such as the first space-based detector, this noise must be larger than the instrumental noise to be detected, and one must have great confidence in the detector in order to claim that the observed noise is external.  This is how the cosmic microwave background was originally discovered in a radio telescope.  

If there are two detectors, then one can multiply the outputs of the two detectors together and sum (integrate).  In this way, the random wave field in one detector acts like a matched filtering template, matching the random field in the other detector.  This allows the detection of noise that is below the instrumental noise of the individual detectors.  For this to work, the two detectors must be close enough together to experience the same random wave field.  In practice, the sensitivity of this method falls off rapidly with separation if the detectors are more than a wavelength apart.

\subsubsection*{Measure of the strength of random gravitational waves}
When describing the strength of a random wave field, it is not appropriate to measure the amplitude of any single component.  Rather, the r.m.s.\ amplitude of the field is the observable quantity.  It is common to use an equivalent measure, the energy density $\rho_{gw}(f)$ in the radiation field as a function of frequency $f$.  For a cosmological field, what is relevant is to normalize this energy density to the critical density $\rho_c$ required to close the universe.  It is thus conventional to define
\begin{equation}\label{eqn:cosdef}
\Omega_{gw}:= \oderiv{\rho_{gw}/\rho_c}{\ln f}.
\end{equation}
This is roughly the fraction of the closure energy density in random gravitational waves between the frequency $f$ and $2.718f$.  

Current and planned detectors may reach a sensitivity of $\Omega_{gw} \sim 10^{-9}$ at 1~mHz and $10^{-10}$ at 40~Hz, but there is a possibility that backgrounds due to other sources (binary white dwarf systems and r-mode spindown, as discussed above) could obscure a cosmological background at these levels.

\subsubsection*{Predicted spectrum of cosmological radiation}
The simplest models of inflation suggest that the spectrum of the gravitational wave background should be flat, so that $\Omega_{gw}$ is independent of frequency over a very large range of frequencies.  In this case, the observed fluctuations in the cosmic microwave background radiation set a limit on gravitational radiation at ultra-low frequencies, and this constrains the energy density in the observable range (0.1~mHz to 10~kHz) to below about $10^{-13}$ of closure.  This will be too small to be seen by the current and planned detectors on the ground or in space.

But there is a great deal of room in these models for other spectra.  The period before inflation may produce initial conditions for the phase of parametric amplification that give large amounts of radiation in the observable frequency range.  One family of models based in superstring theory has a spectrum that rises at high frequencies.  If a cosmological background from inflation or from cosmic defects can be observed, it will contain important clues to the nature of the theory that unifies gravitation with the rest of quantum physics.

\subsection*{Conclusions}
The first few years of the 21st century should see the first direct detections of gravitational radiation and the opening of the field of gravitational wave astronomy.  Beyond that, over a period of a decade or more, one may expect observations to yield important and useful information about binary systems, stellar evolution, neutron stars, black holes, strong gravitational fields, and cosmology.

If gravitational wave astronomy follows the example of other fields, like {\sc X-Ray Astronomy} and {\sc Radio Astronomy}, then at some level of sensitivity it will begin to discover sources that were completely unexpected.  Many scientists think the chance of this happening early is very good, since the processes that produce gravitational waves are so different from those that produce the electromagnetic radiation on which most present knowledge of the universe is based, and since more than 90\% of the matter in the universe is dark and interacts with visible matter only through gravitation.

Present and planned detectors are known not to be ideal for some kinds of gravitational wave sources.  Sensitive measurements of a cosmological background of radiation from the big bang may not be possible with these instruments if the spectrum follows the predictions of ``standard'' inflation theory.  Most of the normal mode oscillations of neutron stars will be very hard to detect, because the radiation is weak and at a high frequency, but the science there is compelling: neutron-star seismology may be the only way to probe the interiors of neutron stars and understand these complex and fascinating objects. Detector technology will continually improve, and these sources provide important long-term goals for this field.  There will clearly be much to do after the first observations are successfully made.

\subsection*{Bibliography}
Folkner, W M, ed., 1998 {\em Laser Interferometer Space Antenna - AIP Conference Proceeding 456} (Woodbury, NY: American Institute of Physics)\\[2mm]
Marck J-A, Lasota J-P, eds., 1996 {\em Relativistic Gravitation and
Gravitational Radiation} (Cambridge and New York: Cambridge University Press)\\[2mm]
Thorne, K S 1994 {\em Black Holes and Time Warps: Einstein's
Outrageous Legacy} (New York: W. W. Norton and Co)\\[2mm]
Wald, R M, ed., 1998 {\em Black Holes and Relativistic Stars} (Chicago: University of Chicago Press)\\[1cm]

\end{document}